# RF modulation studies on the S band pulse compressor

SHU Guan(束冠)[1,2]　　ZHAO Feng-Li(赵凤利)[1)]　　PEI Shi-Lun(裴士伦)[1)]　　XIAO Ou-Zheng(肖欧正)[1)]

[1]Laboratory of Particle Acceleration Physics & Technology, Institute of High Energy Physics,
Chinese Academy of Sciences, Beijing 100049, China

[2]University of Chinese Academy of Science, Beijing 100049, China

**Abstract:** An S band SLED-type pulse compressor has been manufactured by IHEP to challenge the 100 MW maximum input power, which means the output peak power is about 500 MW at the phase reversal time. In order to deal with the RF breakdown problem, the dual side-wall coupling irises model was used. To further improve the reliability at very high power, amplitude modulation and phase modulation with flat-top output were taken into account. The RF modulation studies on an S-band SLED are presented in this paper. Furthermore, a method is developed by using the CST Microwave Studio transient solver to simulate the time response of the pulse compressor, which can be a verification of the modulate theory. In addition, the experimental setup was constructed and the flat-top output is obtained in the low power tests.

## 1　Introduction

The SLAC energy doubler (SLED) type pulse compressors play an important role in the linear accelerators to increase the efficiency of the klystron RF power. An S band SLED-type pulse compressor has been manufactured by IHEP to challenge 100 MW input peak power. At just one compressed pulse time (usually equal to the filling time of the travelling wave accelerating structure) from the incident RF pulse end, the incoming pulse phase is reversed 180° by the PSK (phase shift keying) switcher, the output peak power will reach 500 MW. The extreme high power leads to the sparking phenomena around the SLED coupling irises and the first several accelerating cells. A significant reduction of the electric fields near the irises had been achieved by adopting dual side-wall coupling irises model. High power test results show that the maximum input power can reach 85 MW [1].

To further improve the high power reliability, the amplitude modulation (AM) and the phase modulation (PM) of the SLED were considered to decrease the peak power while the integrated power over the compressed pulse time is tried to keep the same level [2-4]. For AM, the input power is slowly increased to compensate the damped radiation power of the storage cavities during the pulse compressed period. For PM, the incoming RF pulse phase is manipulated while the amplitude is kept constant. RF modulation of the SLED is also an effective way to compensate the beam loading effects in multi-bunch operation [5, 6].

In this paper, we present the AM and PM theory based on the equivalent circuit model of the SLED. Inspired by the innovative idea of the CST Microwave Studio (MWS) [7] transient domain simulation developed by Pohang Accelerator Laboratory (PAL) researchers [8], we propose an analogous method to obtain the flat-top output based on the simulation, which can be a verification of the modulate theory. To further confirm our RF modulation study, the low power experimental setup was constructed, in which the flat-top output is achieved as well.

## 2　Modulation theory

Figure 1 shows the equivalent circuit model of the SLED. The energy storage cavity can be regarded as an oscillating circuit. The voltage source refers to the RF generator.

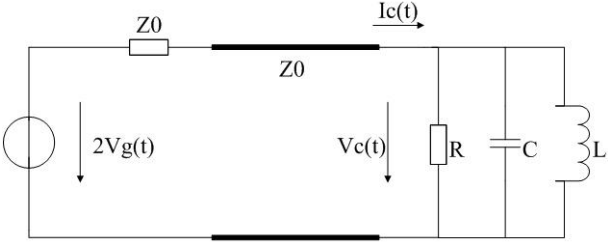

Fig. 1.　Equivalent circuit model of the SLED.

According to the Kirchhoff's law, the following differential equation can be obtained (dots mean derivatives with respect to time) [2].

$$Vr + \tau \dot{Vr} = \Gamma Vg - \tau \dot{Vg}, \quad (1)$$

where Vg is the equivalent complex voltage of the SLED input while Vr is the output. $\tau = 2Q_l/\omega_c$ is the filling time of the storage cavity, $Q_l$ and $\omega_c$ are the loaded Q and the resonant angular frequency. $\Gamma$ is the reflection coefficient and can be defined as $(\beta-1)/(\beta+1)$ with $\beta$ the coupling factor.

## 2.1　AM method

Figure 2 shows the schematic diagram of the AM method, the RF power is fed into the SLED at time $t_0$, the input phase is reversed by 180° with the amplitude dropping to $V_0$ at time $t_1$, and then the incident RF amplitude is modulated (increased continually) to compensate the damped radiation power of the storage cavities until the RF pulse ends at time $t_2$. During $t_1 \leq t < t_2$, the output is the superposition of the input power and the radiation power, the flat-top output can be acquired by an appropriate input waveform.

Conceptually, the Vg experiences a mutation at time $t_0$, $t_1$ and $t_2$, while the equivalent voltage of the cavity radiation field remain unchanged. Therefore, it can be given that

$$\Delta Vr + \Delta Vg = 0, \quad (2)$$

where ΔVr and ΔVg are the transient variation of the Vr and Vg at time $t_0$, $t_1$, $t_2$. The input and output amplitude variations (as shown in Fig. 2) can be calculated by solving the Eq. (1) with the boundary condition Eq. (2). In our case, the specifications of the SLED used in the modulation are summarized in the Table 1. The theoretical value of the integrated power gain over the compressed interval is 2.3.

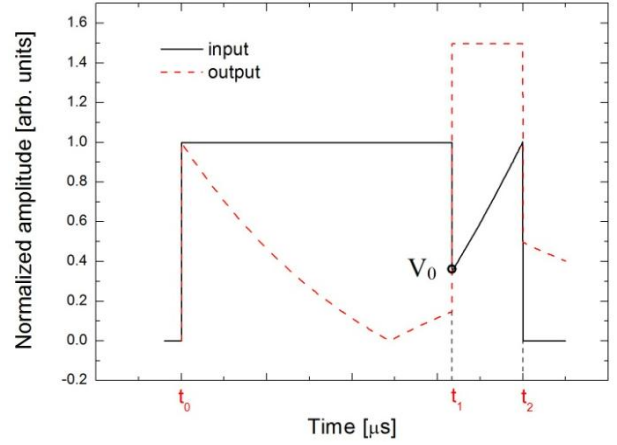

Fig. 2. (color online) Schematic diagram of the AM method.

Table 1.　Main parameters of the SLED.

| Frequency | 2998 | MHz |
|---|---|---|
| Resonant mode | $TE_{0,1,5}$ | |
| Coupling coefficient | ~ 5 | |
| Unload Q factor | ~ 100,000 | |
| Input pulse length | 4 | μs |
| Output pulse length | 0.83 | μs |

Assuming the maximum of the input RF amplitude at the end of the pulse is normalized to 1, the output pulse waveform absolutely depends on the $V_0$ value. Fig. 3 shows the input/output field amplitude with three different $V_0$. The solid and dash lines in the Fig. 3 represent the normalized input and output field amplitude, respectively. The situation of $V_0=1$ corresponds to the original operation mode of the SLED, in which the output has a spiky part due to a phase reversion. When $V_0$ is set to 0.6, the partial flat output can be obtained. Furthermore, the full flat-top output can be obtained when $V_0$ is decreased to 0.3. It can be clearly seen that the peak power at time 3.17μs and the integrated power over the compressed time is reduced by AM process.

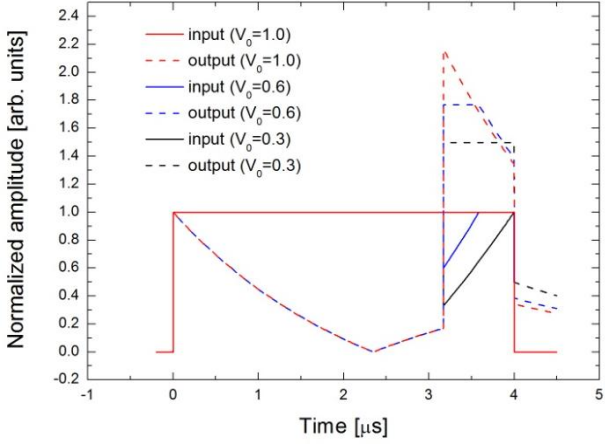

Fig. 3. (color online) Input and output RF amplitude with different $V_0$.

Figure 4 shows the dependence of the SLED power gain on the $V_0$. Here, the output integrated power over the compressed interval is named as the average power. Both the average power and the peak power are reduced by introducing AM process. However, when the $V_0$ decreases from 1 to 0.6, the average power is ~8% lower than the maximal value, while the peak power is decreased dramatically by ~33% of the maximum. The sharply reduced peak power can improve the high power reliability significantly. From the view of the power utilizing efficiency, the partial flat output is more suitable than the full flat-top output [9].

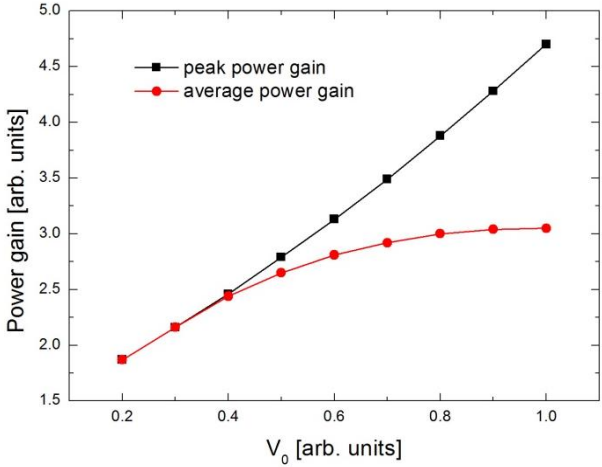

Fig. 4. (color online) Power gain as a function of $V_0$.

Due to the non-linear input/output characteristics of the klystron at high power, AM performance in the klystron saturation regime will be taken into account in detail.

### 2.2 PM method

Assuming the RF power is fed into the SLED at time $t_0$, the input has a phase jump with a step $\varphi_0$ (generally much less than $180^\circ$) at time $t_1$, then the phase is increased continuously until the RF pulse ends at $t_2$. The phase modulation based upon the differential Eq. (1) is carried out during $t_1 \leq t < t_2$, then a compressed pulse with constant amplitude can be acquired. The detail formula derivation can be found in ref. [2]. As is known to all, there is a large phase variation of the output during the compressed interval. Fig. 5 shows the dependence of the average power gain and the output phase variation on the phase jump step $\varphi_0$. The average power gain is proportional to $\varphi_0$, while the output phase variation increases with $\varphi_0$ as well. For the time duration $t_1 \leq t < t_2$, the output phase experiences a large variation. The compressed pulse will be fed into the accelerator structure, the large phase variation will lead to degradation of the beam performance.

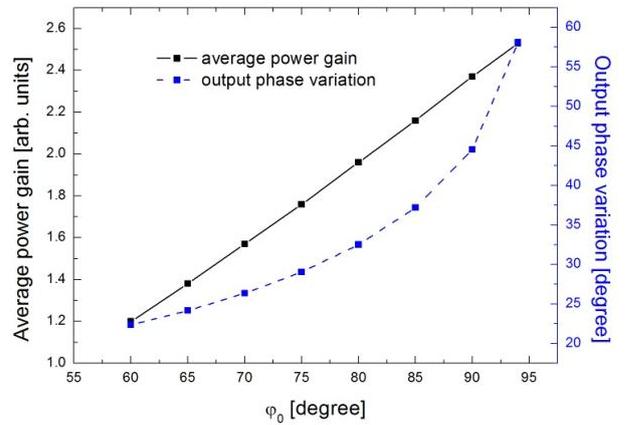

Fig. 5. (color online) Average power gain and output phase variation depending on the phase jump step $\varphi_0$.

In order to reduce the output phase variation, the RF generator frequency can be set at a relatively higher value (e.g. 150 kHz) than the cavity resonant frequency. For this scenario, Eq. (1) can now be modified as follows

$$V_r(1 + j2\pi\tau\Delta f) + \tau\dot{V}_r = (\Gamma - j2\pi\tau\Delta f)V_g - \tau\dot{V}_g, \quad (3)$$

where $\Delta f = f_0 - f_c$ is the frequency shift, $f_0$ the driven frequency, $f_c$ the resonant frequency of the cavity.

The input and output amplitude and phase variations can be calculated by solving the Eq. (3). The specifications of the phase modulated SLED are listed in Table 1. Fig. 6 shows the input and output amplitude and phase shapes. At time 3.17 μs, an input RF phase jump of 94° is introduced, the average power gain with constant output is 2.36. By comparing Fig. 5 and Fig. 6, the output phase variation decreases from several tens of degree to several degree.

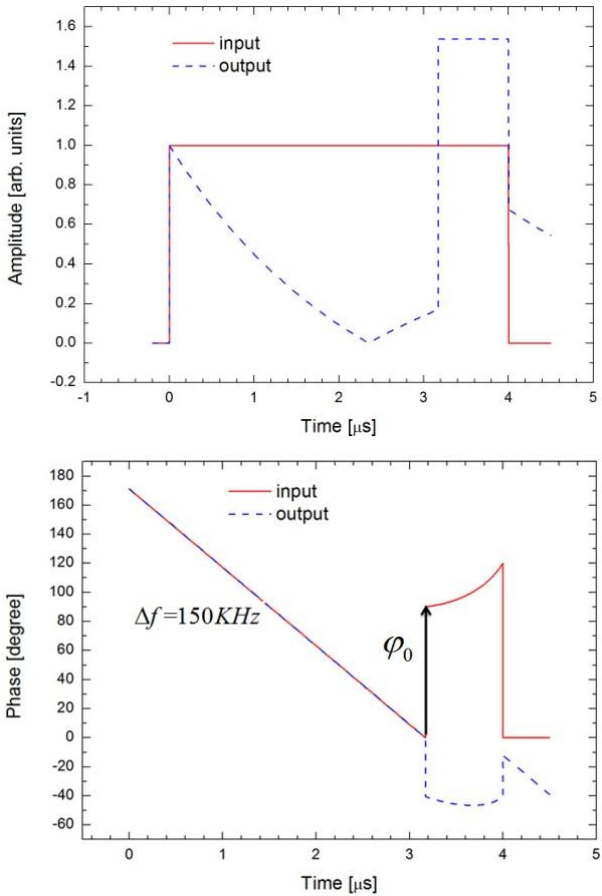

Fig. 6. (color online) Input and output RF amplitude (upper) and phase (lower) variation.

The average power gain is proportional to $\varphi_0$, once the $\varphi_0$ is determined, an optimal value of the frequency shift can be found to minimize the output phase variation, as shown in Fig. 7. Partial flat output can also be acquired by PM method [10, 11], the output waveform and the power gain is similar to the Fig. 3 and Fig. 4.

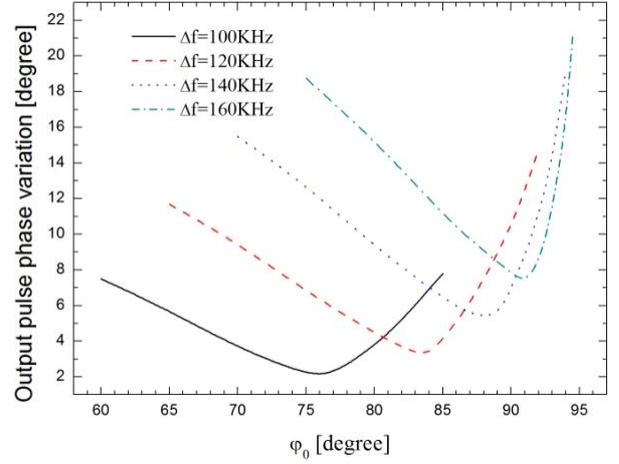

Fig. 7. (color online) Relationship between the output phase variation and the phase jump step $\varphi_0$.

## 3  Transient simulations

By using the MWS transient solver, the SLED response in time domain can be studied qualitatively, then the theoretical study results of the AM and the PM can be confirmed. In the case of the AM process, the input pulse is expressed by $V(t)\sin(2\pi f_0 t)$, and $V(t)$ is the modulated incident RF amplitude function in Fig. 2. By importing the driven signal shown in Fig. 9 (a) into MWS, the SLED response can be simulated, which is shown in Fig. 9 (b).

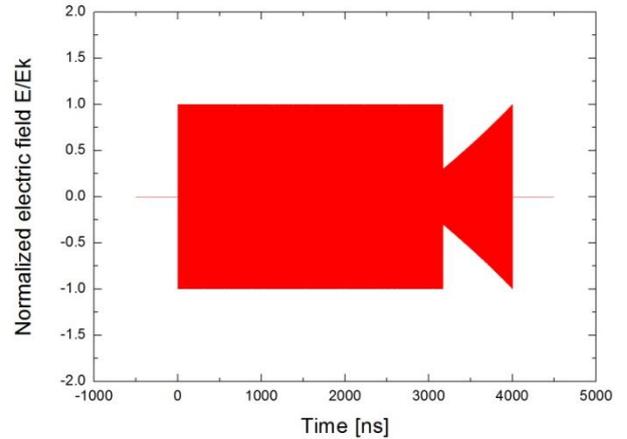

(a)

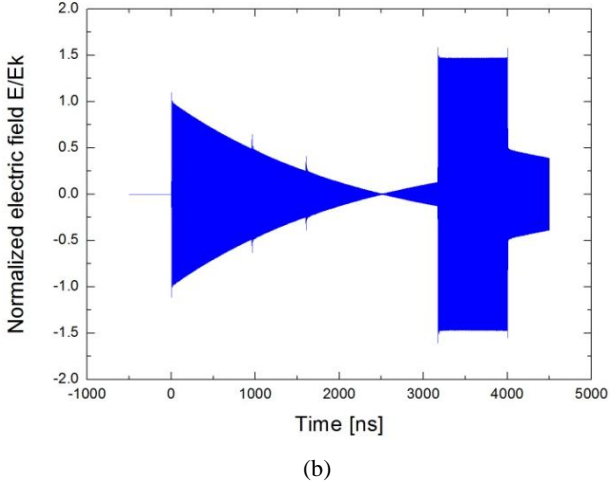

(b)

Fig. 9. (color online) (a) and (b) correspond to the driven and the response signals of SLED simulated by MWS transient solver with AM method.

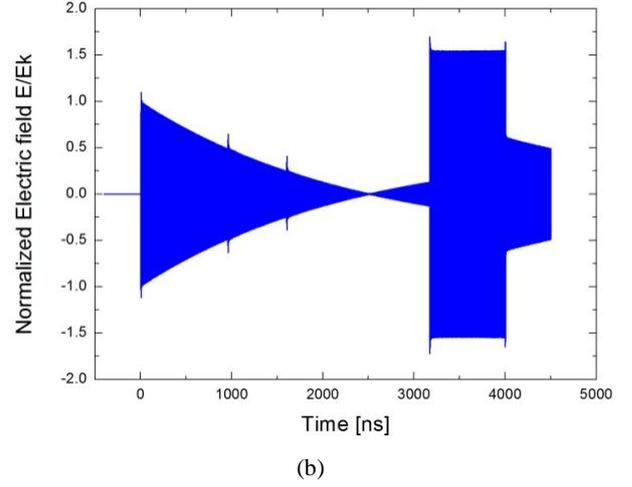

(b)

Fig. 10. (color online) (a) and (b) correspond to the driven and the response signals of SLED simulated by MWS transient solver with PM method.

For the PM process, the input pulse is expressed by $\sin(2\pi f_0 t + \varphi(t))$, $f_0$ is set 150kHz higher than the cavity resonant frequency and $\varphi(t)$ is the modulated input phase function in Fig. 7. The driven signal and the SLED response signal simulated by MWS are shown in Fig. 10. The flat-top output is acquired by both AM and PM in the simulation, so this proves the correctness of the theoretical calculations.

## 4 Experiment

To further confirm our RF modulation study, the low power experimental setup was constructed shown in Fig. 11. The carrier wave coming from the RF pulse signal generator is modulated by I and Q control levels which are generated by two arbitrary waveform generators. Then the modulated pulse is fed into the SLED cavities. The peak power meter is used to monitor the output.

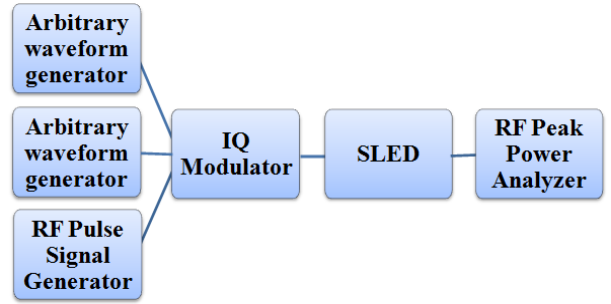

Fig. 11. (color online) The schematic layout of the test.

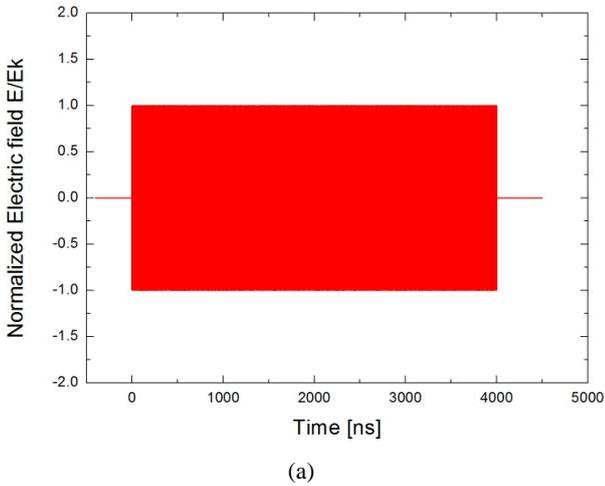

(a)

The parameters of the SLED used in the experiment are given in Table 1. At first, the frequency of the SLED cavities was tuned to $f_0 = 2998$ MHz, then the AM and PM processes were implemented. Fig. 12 shows the cold test results. In the case of the AM, the value of $V_0$ was set to 0.3 at $3.17\,\mu s$, the flat-top output was obtained as shown in Fig. 12 (a). The output average power was 2.33 times the input and the flatness was better than 95%. Fig. 12 (b) corresponds to the PM method, the RF generator frequency was set as 2998.150MHz (150 kHz higher than the resonant frequency), an input RF phase jump of 94° is adopted at time 3.17 μs. The flat-top output was obtained with an average power gain of 2.29 and the flatness was better than 95%.

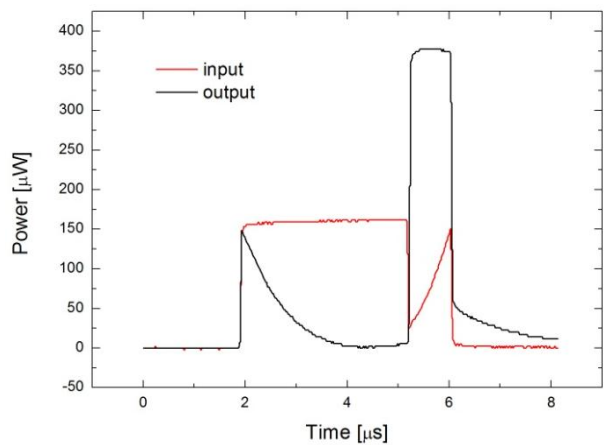

(a)

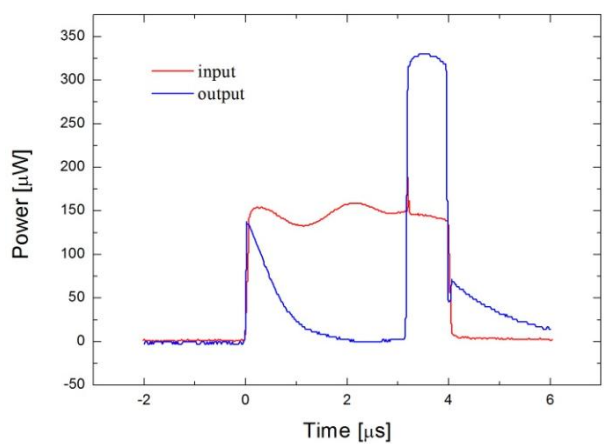

(b)

Fig. 12. (color online) (a) and (b) correspond to the power measurement of the SLED using AM and PM.

## 5 Conclusions

An excessive surface field within the cavity leads to potentially serious breakdown problems. The maximum RF input power of the SLED-type pulse compressors can be enhanced by introducing the RF modulation. We perform the AM and the PM theoretical analysis, the flat-top output was obtained both in the MWS simulation and the low power test. As the result of the test, the average power gain of the two modulate methods are almost the same. High power test will be concerned in the future.